\documentclass[amssymb,prb,aps,showpacs,twocolumn,showpacs]{revtex4}
\usepackage{graphicx}
\usepackage{epstopdf}
\begin{document}

\title{\textbf{Entanglement manipulation by a magnetic pulse
in $Gd_3N@C_{80}$ endohedral metallofullerenes on a $Cu(001)$ surface }}
\author{Oleg V.Farberovich$^{1,2,3}$, Vyacheslav S.Gritzaenko$^{2}$  }
\affiliation{
$^1$School of Physics and Astronomy, Beverly and Raymond Sackler Faculty
of Exact Sciences,
Tel Aviv University, Tel Aviv 69978, Israel\\
$^2$International Center "Smart Materials", Southern Federal University,
Zorge 5, 344090, Rostov-on-Don,Russia\\
$^3$Voronezh State University, Voronezh 394000, Russia\\}
\begin{abstract}
In this paper we present result of theoretical calculation of entanglement within a spin structure of $Gd_3N@C_{80}$ under the influence of rectangular impulses. Research is conducted using general spin Hamiltonian within SSNQ (spin system of N-qubits). Calculation of entanglement with variable impulse is performed using the time-dependent Landau-Lifshitz-Gilbert equation with spin-spin correlation function. We show that long rectangular impulse (t=850ps) can be used for maintaining of entanglement value. This allows us to offer a new algorithm which can be used to reduce the challenge of decoherence to logical scheme optimization.
\end{abstract}
\maketitle
\date{\today}

\section{introduction}
Quantum entanglement plays a central role in quantum
information and quantum computation.\cite{1,2}
 Though the theoretical investigations of the magnetism and
electronic structure of lanthanides have been performed decades
ago at the same time  fundamental results of the time dependent spin structure appear under a new light
how a molecular spintronics.\cite{3}
 Quantum entanglement (QE) also exists on the
nanoscale in solid state and molecular systems.\cite{4} Condensed matter systems are
especially interesting subjects for studying entanglement, f.e. entanglement
of spins, for several reasons. First and foremost,
condensed matter is the basis of a sizeable portion of
todays research and development in spintronics and in-
formation industries. Understanding QE in solids and
nanostructures could open a complete new perspective
on those research directions. Moreover, the source of QE
of spin states is the exchange interaction, which is known to
be strong in condensed matter systems.\cite{5}
In the case of solids and molecules, however, detecting QE can be tricky.

Key challenges in
building a quantum computer from spin qubits in physical
systems are preparation of arbitrary spin states, implementation
of the arbitrary qubit evolution, reading
out the qubit states, overcoming of the decoherence and
doing all this on a large scale; that is, with a large number
of qubits or a spin system of N-qubits (SSNQ) with
the definite spin structure.\cite{5} For the purposes of implementing
quantum computation, the physical system can
be treated as a SSNQ in which the couplings between
the qubits can be controlled externally. The concept of
a SSNQ can be related to the problem of the quantum
spin structure, where the nontrivial applications may exist
for computers with a limited number N of a qubits.
The precise relationship between the type of the entanglement
and the distribution of the coupling strengths in
the SSNQ can be strongly dependent on external parameters,
such as applied magnetic pulse fields.
In this context, systematic studies of the relationship between
the quantity and nature of the entanglement and
the spin structure of the SSNQ via magnetic pulse has been pursued in order
to identify the optimal spin structures to create specific
types of the entanglements.
Here we  use the spin correlating states\cite{6} to study entanglement
in metallic clusters  of Gd trimers in $Gd_3N@C_{80}$
endohedral metallofullerene on a Cu(001) surface\cite{7}.

The relevance of
the $Gd_3N@C_{80}$ molecular magnet under a magnetic pulses to
be considered for qubits perspectives will be discussed in
this paper. For this purpose the time-dependent spin
dynamics is involved within a micromagnetic
model based on the general spin Hamiltonian (GSH) formalism,
where the spin is considered as a spatial  and time-dependent
continuous function. In this case the dynamics
of spins obeys the Landau-Lifshitz-Gilbert (LLG)
equation\cite{8}, involving the various energy contributions from
the exchange, magnetostatic and Zeeman interactions,
and magnetic anisotropy of the spin system. It is
known, that correlations are of a great importance in the
study of spin systems. They are directly related to
the entanglement between different atomic spins, which
can be employed in the field of quantum information
processing. Therefore entanglement arising in the correlated
quantum spin systems has become one of the most
widely investigated phenomenon in quantum physics during
last years. Now it is well known how to create
entanglement in quantum spin systems, but how it propagates
inside a system is still an important fundamental
question in quantum information theory. Therefore the
theoretical descriptions of the time-dependent properties
of the correlated quantum spin systems is very important
nowdays. Recently it was claimed, that the controlled
and manipulated entanglement in the quantum
spin systems could be realized precisely and effectively
by means of the required dynamical operations in presence
of the magnetic pulse.\cite{9} Moreover, it was shown
also that the solid state spin-based system with a definite
spin structure can be considered as a spin system
of N-qubits, which demonstrates the prolonged
time of the decogerence demanded for quantum information
processing. For implementation of quantum
computation one can treat a quantum spin system as
a SSNQ, where the couplings between the spin qubits
can be controlled externally, for instance by the applied
magnetic pulse. Therefore the systematic
studies of inherent relationship between the strength of
the entanglement and the peculiarities of spin structure
of a SSNQ are carried out in order to find the optimal
time- dependent spin structures with specific types of
the controlled and engineered entanglements.
\section{The theoretical approach}
\subsection{The spin-dynamics simulations}
The dynamic behavior of a spin is determined by the equation
of motion, which
can be derived from the quantum theory with the spin Hamiltonian
$\widehat H_{spin}$ that calculated the spin structure of a magnetic cluster with
Hamiltonian
\begin{equation}\label{GSH}                                                                                                                 \widehat H_{spin} = \widehat H_{ex} + \widehat H_{an} +
\widehat H_{ZEE} + \widehat H_a(t).
\end{equation}                                                                                                                   The first term $\hat H_{ex}$ is the Heisenberg Hamiltonian, which represents the isotropic
exchange interaction, $H_{an}$ is the exchange Hamiltonian the term due to the
axial single-ion anisotropy, and $H_{ZEE}$ is the interaction between the
spin system and the external magnetic field.
This Hamiltonian, which describes
the interaction of the spin
$\widehat{\bf S}$ with the external
magnetic field, given by its flux ${\bf H}_{eff}$, can be expressed as:
\begin{equation}
\widehat H_{spin}=-{\bf H}_{eff}\widehat{\bf S}
\end{equation}
where the effective magnetic field ${\bf H}_{eff}$ is
an external magnetic
field ${\bf H}_z$, the anisotropy
fields ${\bf H}_{an}$, the exchange interaction ${\bf H}_{ex}$
and external magnetic pulse field ${\bf H}_{pulse}(t)$.
Here we use the approximation for ${\bf H}_{eff}
\Rightarrow{\bf H}_{eff}^{mean}$
with replacement $\widehat{\bf S}\Rightarrow{\bf M}_s=\gamma\langle\widehat
{\bf S}\rangle$.
Use \cite{W26} we obtain that
\begin{equation}
\frac{\partial\langle\widehat {\bf S}\rangle}{\partial t}=\frac{1}{1+\lambda^2}
\langle\widehat {\bf S}\rangle\times{\bf H}^{mean}_{eff}-
\frac{\lambda}{1+\lambda^2}\langle\widehat{\bf S}\rangle\times
(\langle\widehat{\bf S}\rangle\times{\bf H}^{mean}_{eff})
\end{equation}
The effective magnetic field ${\bf H}_{eff}^{mean}$ is given by the
free magnetic energy
variational with magnetization:
$$                                                                                                                                                               {\bf H}_{eff}^{mean}({\bf M}_s,t) =-\frac{\delta F}{\delta{\bf M}_s},
$$                                                                                                                                                                where $F$ is the free energy of the magnetic nanosystem.                                                                                               
The effective field ${\bf H}_{eff}^{mean}$ can be derived from
the free energy functional
\begin{equation}
{\bf H}_{eff}^{mean}=-\frac{\delta (F(M_s,H_z)+F(t))}{\delta{\bf M}_s}=
-\frac{\partial F(M_s,H_z)}{\partial{\bf M}_s}+H_{pulse}^x(t),
\end{equation}
We have derived a general form of the time-dependend spin equation for a system of the spins                                                                             precessing in an effective magnetic field with specifying the interactions
in the magnetic cluster on a surface.

A spin structure is defined only proceeding from the spin model of a cluster.
Here we use to calculate a spin structure by the ITO method within the generalized spin
Hamiltonian $\widehat H_{spin}$.

Since in  further researches the anisotropic part of a cluster will be only scalar,
the magnetic properties of the anisotropic system do not depend on the
direction of the magnetic field.
Thus we can consider the external magnetic
field $H_z$ directed along arbitrary axis
$\it z$ of the cluster coordinate frame that is
chosen as a spin quantization axis. In
this case the energies of the system will be
$\epsilon_{\mu}(M_s) +   g_e M_s H_z$, where
$\epsilon_{\mu}(M_s)$
are the eigenvalues of the spin-Hamiltonian containing
the magnetic exchange and
the double exchange contributions
(index $\mu$ runs over the energy levels with given
total spin protection ${\bf M}_s$).

We use them further to define the spin-spin correlation functions
$\mathcal C_{ij}$ for the entangled ground state $\mid\it
SM^{(ij)}_0\rangle = \alpha_i\mid\it SM\rangle_i+
\beta_j\mid\it SM\rangle_j$ of the system \cite{corr}:

\begin{equation}
\mathcal C_{ij}^{\alpha \beta}(t)=\langle\it SM^{(ij)}_0\mid\widehat
{S}_i^\alpha(t)\widehat {S}_j^\beta(t)\mid\it SM^{(ij)}_0\rangle=
\langle\it S_i^\alpha\rangle(t)\langle\it S_j^\beta \rangle(t),
\end{equation}
where $\alpha, \beta \in x, y, z$.

The ground state is in the subspace $\mathcal {A}$ for which
$M^{(\mu)}$ = 0 for all $\mu$.

The entanglement entropy between a subspace $\mathcal {A}$ and
the rest of the system $\mathcal {R}$ is given by:

\begin{equation}
S_\mathcal{A}=-\rm{Tr}(\rho_\mathcal{A}\log_2\rho_\mathcal{A}),
\end{equation}

where $\rho_\mathcal {A}$ is the reduced density matrix of the
subspace $\mathcal {A}$ obtained by tracing out over all those parts of
the Hilbert space not associated with $\mathcal {A}$. We consider the
subspace $\mathcal {A}_{ij}\equiv \lbrace i,j\rbrace$ consisting of all
spin pairs (not only neighboring) $i$ and $j$ of the ground states
$\mid\it S^{(\mu)}M^{(\mu)}=0\rangle$ \cite{171}. For the rest of the
system $\mathcal {R}$ is given by $\mathcal {R}_{kl}\equiv \lbrace
k,l\rbrace$. The matrix elements of the reduced density matrix, needed to
calculate the entanglement, can be written in terms of the spin-spin
correlation functions $\mathcal C_{ij}^{\alpha\beta}(t)$ (see eq. (5)):

\begin{equation}
\rho_\mathcal {A}^{(ij)}(t)=\sum_{\alpha,\beta\in{x,y,z}}
\langle\it SM^{(ij)}_0\mid\widehat {S}_i^\alpha(t)\widehat
{S}_j^\beta(t)\mid\it SM^{(ij)}_0\rangle
\rho_{\mu\nu}^{(ij)},
\end{equation}

where
$$
\rho_{\mu\nu}^{(ij)}=\mid\it SM^{(ij)}_0\rangle_\mu\otimes\mid\it
SM^{(ij)}_0\rangle_\nu.
$$

Suppose $\lbrace\mid\mathcal {A}_{ij}\rangle\rbrace$ and
$\lbrace\mid\mathcal {R}_{kl}\rangle\rbrace$ are the orthonormal basis
states of the many-body Hilbert space of the subsystems $\mathcal {A}$
and $\mathcal {R}$. A general quantum spin state of the composite system
can be described by � wave function \cite{Rad}:

\begin{equation}
\mid\Psi\rangle=\sum_{ij,kl}C_{ij,kl}\mid\mathcal
{A}_{ij}\rangle\mid\mathcal {R}_{kl}\rangle.
\end{equation}

Here, a rectangular matrix $\bf {C}$ can be presented always  in the form
$\bf {UDV^{\dagger}}$, where $\bf {U}$ is unitary, $\bf {D}$ is diagonal
and the rows of $\bf {V}$ are orthonormal. It is known already as  the
singular-value decomposition (SVD).
Using this decomposition in eq. (8) and forming a new basis by combining the
$\mid\mathcal {A}_{ij}\rangle$  with $\bf {U}$ and the $\mid\mathcal
{R}_{kl}\rangle$ with $\bf {V^{\dagger}}$, one can obtain the Schmidt
decomposition \cite{13}:

\begin{equation}
\mid\Psi\rangle=\sum_{k=1}^{Rank}\sigma_k\mid\Phi_k^{\mathcal
{A}}\rangle\mid\Phi_k^{\mathcal {R}}\rangle,
\end{equation}

which represents  the total wave function $\mid\Psi\rangle$ of the system
as a single sum of products of the orthonormal functions. Here the $Rank$
number of the terms is limited by the smallest one of the two Hilbert
spaces and the weight factors $\sigma_k$ are the matrix elements of the
diagonal matrix $\bf {D}$.
\begin{figure}[h!]
\begin{center}
\includegraphics[width=1.0\columnwidth]{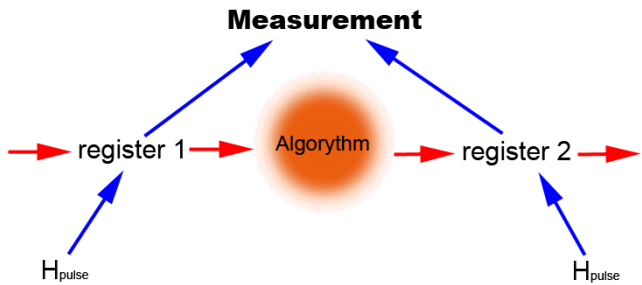}
\end{center}
\caption{ Principle of "reload"}
\label{fig.1}
\end{figure}
If $\mid\Psi\rangle$ is normalized, their
absolute magnitudes squared sum to one. The entanglement properties of a
system are performed with the set of $\sigma_k$:

\begin{equation}
S_\mathcal {A}(t)=-\sum_{k=1}^{Rank}\tau_k(t)\log_2\tau_k(t),
\end{equation}

where

\begin{equation}
\tau_k(t)=\langle\it S_i^\alpha\rangle(t)\langle\it
S_j^\beta\rangle(t)\sigma_k^{\rm 2}.
\end{equation}

\section{results and discussions}
Quantum entanglement is primary criteria of quantum computer efficiency. Due to that fact retention of entanglement for a long time is main challenge facing researchers. This problem calls decoherence. A great number of increase the coherence time methods for circumvent this phenomenon are invented. But all of these algorithms cannot completely eliminate decoherence.

\begin{figure}[h!]
\begin{center}
\includegraphics[height=9cm, width=8.5cm]{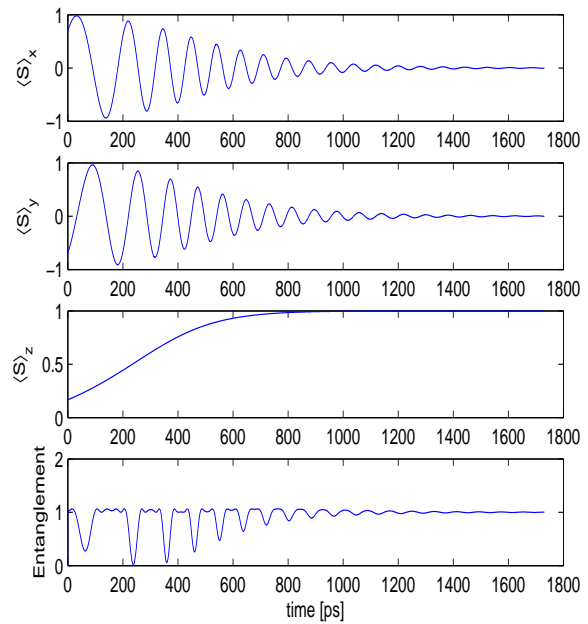}
\end{center}
\caption{(Color online) The evolution of the time-dependent quantum
mechanical expectations $\langle {S}^{x,y,z}\rangle(t)$ of the $Gd_3N@C_{80}$
molecule under  magnetic pulses H$_{pulse}$=0T.}\label{fig.2}
\end{figure}
\begin{figure}[h!]
\begin{center}
\includegraphics[width=1\columnwidth]{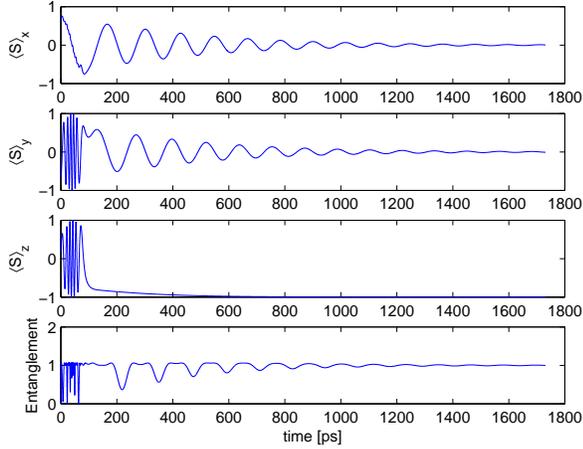}
\end{center}
\caption{(Color online) The evolution of the time-dependent quantum
mechanical expectations $\langle {S}^{x,y,z}\rangle(t)$ of the $Gd_3N@C_{80}$
molecule under  magnetic Gauss pulses H$_{pulse}$=5.3T.}\label{fig.3}
\end{figure}

We offer a new method in which entanglement maintained by two long rectangular pulses. It allows us to hold maximal measure of quantum entanglement until pulse affect on system. Within a framework of this method we create an algorithm of quantum computer efficiency maintenance that operate under a principle of "reload" (Fig. \ref{fig.1}). At first step by the long rectangular pulse we create entanglement and maintain maximal value of it. During the pulse wave functions of system is measured. At the second step we impact on system by operators of scheme. At the third step of algorithm we do measurement of system wave functions and create a maximal entanglement by the second long rectangular pulse. Due to the fact that this algorithm can be used as a cycle entanglement is maintained on maximal value, it changes only in second step. The priority in this method is not coherence during the quantum computer working time but coherence within work of scheme. This allows us to reduce the challenge of decoherence to scheme optimization which is a solvable problem.
\begin{figure}[h!]
\begin{center}
\includegraphics[width=1\columnwidth]{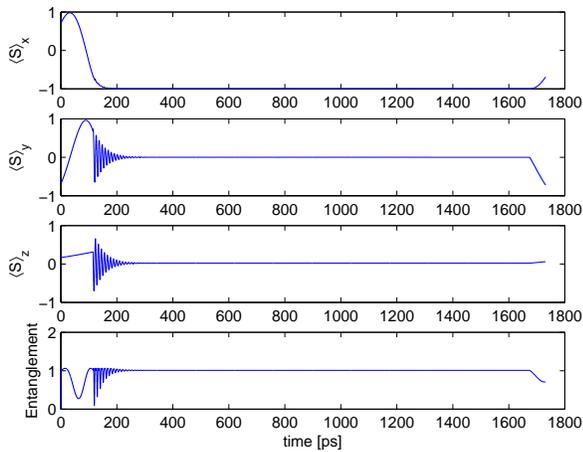}
\end{center}
\caption{(Color online) The evolution of the time-dependent quantum
mechanical expectations $\langle {S}^{x,y,z}\rangle(t)$ of the $Gd_3N@C_{80}$
molecule under  magnetic rectangular pulses H$_{pulse}$=5.3T.}\label{fig.4}
\end{figure}
We studied endohedral fullerene $Gd_3N@C_{80}$ due to the antiferromagnetic coupling between the encapsulated magnetic moments and the electrons situated on the carbon cage. This property makes this compound promising candidate for molecular spintronics.
\begin{center}
\begin{figure}[h!]
\includegraphics[width=1\columnwidth]{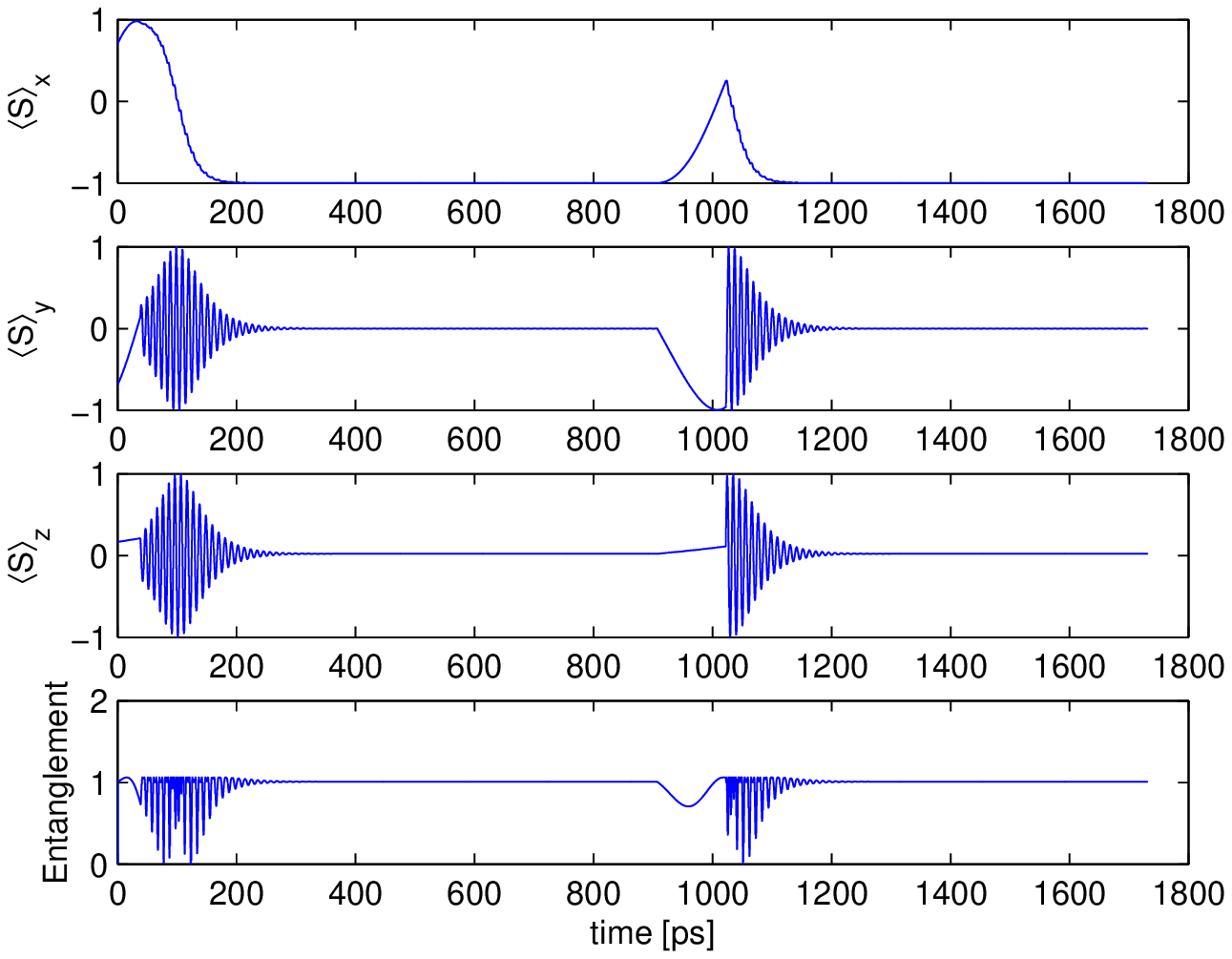}
\caption{(Color online) The evolution of the time-dependent quantum
mechanical expectations $\langle {S}^{x,y,z}\rangle(t)$ of the $Gd_3N@C_{80}$
molecule under two magnetic rectangular pulses H$_{pulse}$=5.3T.}\label{fig.5}
\end{figure}

\begin{figure}[h!]
\includegraphics[width=1\columnwidth]{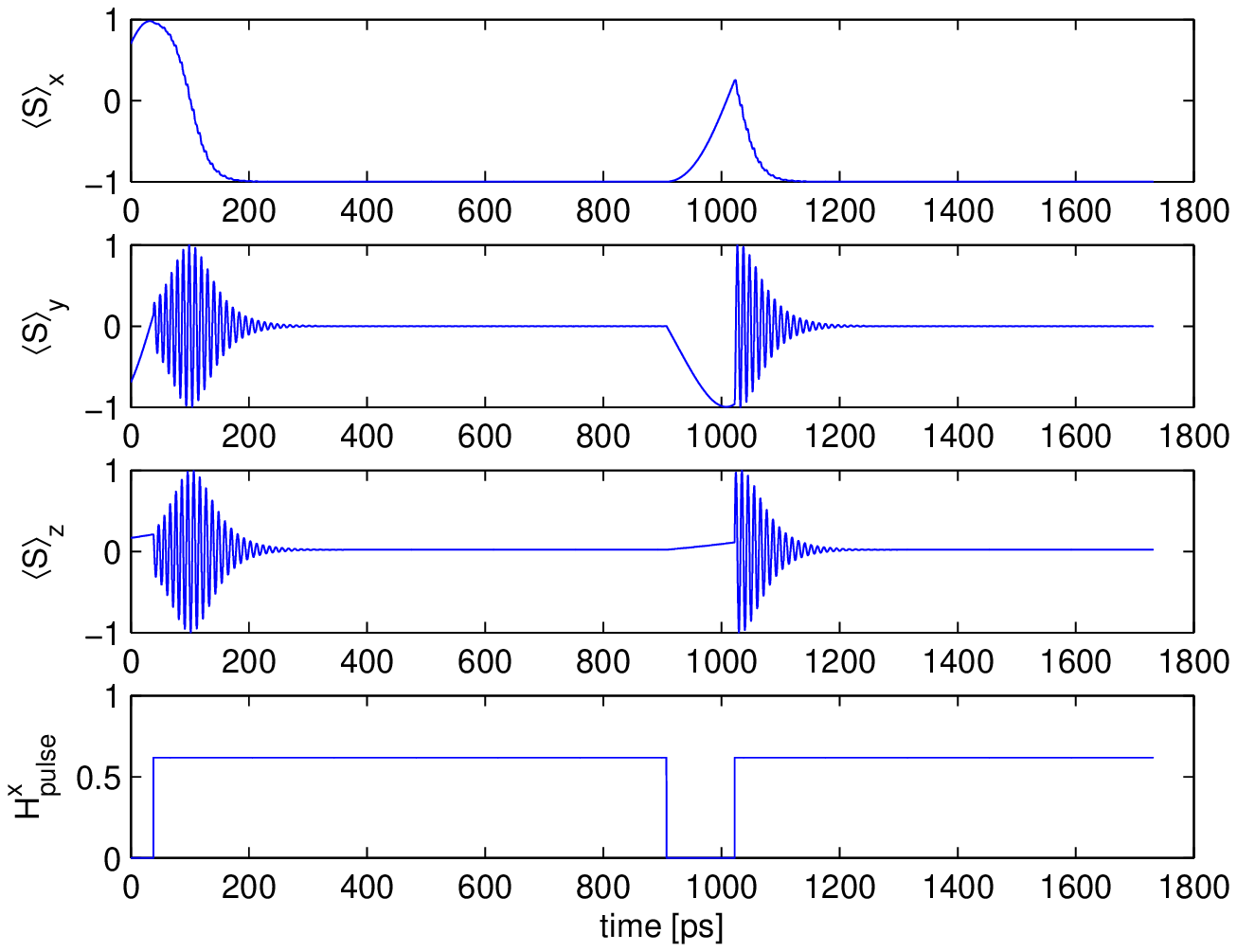}
\caption{(Color online) The evolution of the time-dependent quantum
mechanical expectations $\langle {S}^{x,y,z}\rangle(t)$ of the $Gd_3N@C_{80}$
molecule under two magnetic rectangular pulses H$_{pulse}$=5.3T.}\label{fig.6}
\end{figure}
\end{center}
Entanglement of system without impulse impact (Fig. \ref{fig.2}) is unstable. That’s why probability of decoherence during computing is high. Short rectangular impulse (H$_{pulse}$=5.3T, t=200ps) is the cause of spin-switching (Fig. \ref{fig.3}). Due to that  effect (which is reviewed by Farberocich O.V.) it is possible to create quantum gates and logical schemes. Impact of long rectangular impulse (H$_{pulse}$=5.3T, t=1400ps) on system leads to very interesting results. During this impulse entanglement was maintained on maximal value. This behavior of quantum system shows us that long rectangular impulse can increase a coherence time (Fig. \ref{fig.4}). Results of two long rectangular impulses (H$_{pulse}$=5.3T) impact shown on Fig. \ref{fig.5} (influence of impulses at system) and \ref{fig.6} (long rectangular impulses duration). Value of entanglement is maintained on maximum level during first and second impulses (t=850ps) and  decrease only between them. We offer to impact to system by logical scheme in this moment.
\section{Conclusion}
In present paper we consider a results of entanglement of endohedral fullerene $Gd_3N@C_{80}$ theoretical analysis. Calculation performed with exchange integrals from paper\cite{7}. We show that long rectangular impulse (H$_{pulse}$=5.3T) with duration 850ps can be used for maintaining of entanglement on maximal value. Due to this fact it has become possible to reduce problem of decoherence in quantum computers to challenge of logical scheme optimization by using a two long rectangular impulses with H$_{pulse}$=5.3T and t=850ps each.
\begin{acknowledgements}
The part of this research has been supported by Russian Ministry of Science, Grant RFMEFI 58714X0002.
\end{acknowledgements}

\end{document}